\documentclass{article}
\usepackage{spconf,amsmath,graphicx}
\usepackage{amssymb}
\usepackage{enumitem}
\setlist{nosep, leftmargin=14pt}
\usepackage{siunitx}
\usepackage{caption}
\usepackage[dvipsnames]{xcolor}
\usepackage{booktabs}

\usepackage[colorlinks=true,linkcolor=blue, citecolor=blue, urlcolor=blue]{hyperref}
\usepackage{url}
\usepackage[pagewise]{lineno}

\usepackage{mwe} 

\title{DEMIST: \underline{DE}coupled \underline{M}ulti-stream latent d\underline{I}ffusion for Quantitative Myelin Map \underline{S}yn\underline{T}hesis}

\name{
\begin{tabular}{c}
Jiacheng Wang$^{\star}$, Hao Li$^{\star}$, Xing Yao$^{\star}$, Ahmad Toubasi$^{\dagger}$, Taegan Vinarsky$^{\dagger}$, Caroline Gheen$^{\dagger}$,\\
Joy Derwenskus$^{\dagger}$, Chaoyang Jin$^{\dagger}$, Richard Dortch$^{\ddagger}$, Junzhong Xu$^{\dagger}$, Francesca Bagnato$^{\dagger}$, Ipek Oguz$^{\star}$
\end{tabular}
}
\address{$^{\star}$ Vanderbilt University
$^{\dagger}$ Vanderbilt University Medical Center
$^{\ddagger}$ Barrow Neurological Institute
}
\begin{document}
%
\maketitle

\begin{abstract}
Quantitative magnetization transfer (qMT) imaging provides myelin-sensitive biomarkers, such as the pool size ratio (PSR), which is valuable for multiple sclerosis (MS) assessment. However, qMT requires specialized 20-30 minute scans. We propose DEMIST to synthesize PSR maps from standard T1w and FLAIR images using a 3D latent diffusion model with three complementary conditioning mechanisms. Our approach has two stages: first, we train separate autoencoders for PSR and anatomical images to learn aligned latent representations. Second, we train a conditional diffusion model in this latent space on top of a frozen diffusion foundation backbone. Conditioning is decoupled into: (i) \textbf{semantic} tokens via cross-attention, (ii) \textbf{spatial} per-scale residual hints via a 3D ControlNet branch, and (iii) \textbf{adaptive} LoRA-modulated attention. We include edge-aware loss terms to preserve lesion boundaries and alignment losses to maintain quantitative consistency, while keeping the number of trainable parameters low and retaining the inductive bias of the pretrained model. We evaluate on 163 scans from 99 subjects using 5-fold cross-validation. Our method outperforms VAE, GAN and diffusion baselines on multiple metrics, producing sharper boundaries and better quantitative agreement with ground truth. Our code is publicly available at \url{https://github.com/MedICL-VU/MS-Synthesis-3DcLDM}. 
\end{abstract}


\section{Introduction}
\label{sec:intro}
Quantitative magnetization transfer (qMT) imaging can measure myelin content through biomarkers like the pool size ratio (PSR). PSR is highly specific to demyelination in multiple sclerosis (MS) \cite{bagnato2018selective}. However, qMT acquisitions are lengthy, motion-sensitive, and not routinely available, motivating PSR synthesis from ubiquitous T1w and FLAIR scans. Prior work \cite{sisco2022predicting} suggests PSR maps are partly predictable from clinical images, showing feasibility but also underscoring the need for stronger generative modeling.


Existing synthesis methods based on GANs and UNets \cite{nie2017medical,wang2023self} can produce unrealistic features and lose anatomical detail. Diffusion models like DDPM and DDIM \cite{ho2020denoising,song2020denoising} are more stable and produce better results, but applying them directly in pixel space is too computationally expensive for 3D medical images. 
Latent diffusion models (LDMs) solve this problem by working in the compressed latent space of an autoencoder rather than directly on images, allowing them to generate 3D brain MRI and translate between modalities \cite{rombach2022high,kim2024adaptive,pinaya2023monai_generative,lyu2022conversion}.

Diffusion-based image translation can be conditioned through concatenation, cross-attention \cite{rombach2022high}, spatial adaptive normalization \cite{kim2024adaptive}, or control branches like ControlNet \cite{zhang2023adding}. ControlNet is particularly effective because it freezes the pretrained diffusion backbone and adds trainable parallel branches through zero-initialized connections, preserving learned priors while enabling task-specific control, such as segmentation-guided generation and cross-modality translation \cite{konz2024anatomically,yu2025adaptive,sargood2025cocolit}. Beyond conditioning mechanisms, low-rank adapters (LoRA) \cite{hu2022lora} provide parameter-efficient modulation of attention projections. 

We propose DEMIST: 3D LDM uses both ControlNet and cross-attention for PSR synthesis (Fig.~\ref{fig:architecture}). 
The key idea is to decouple anatomical structure preservation (via ControlNet) from quantitative value generation (via cross-attention), while adapting a frozen diffusion foundation model with LoRA so that task-specific calibration does not compromise pretrained priors.
We found that explicitly separating these objectives improves results because they require different types of conditioning information. Our main contributions are:

\begin{figure*}[ht]
    \centering
    \includegraphics[width=1\linewidth]{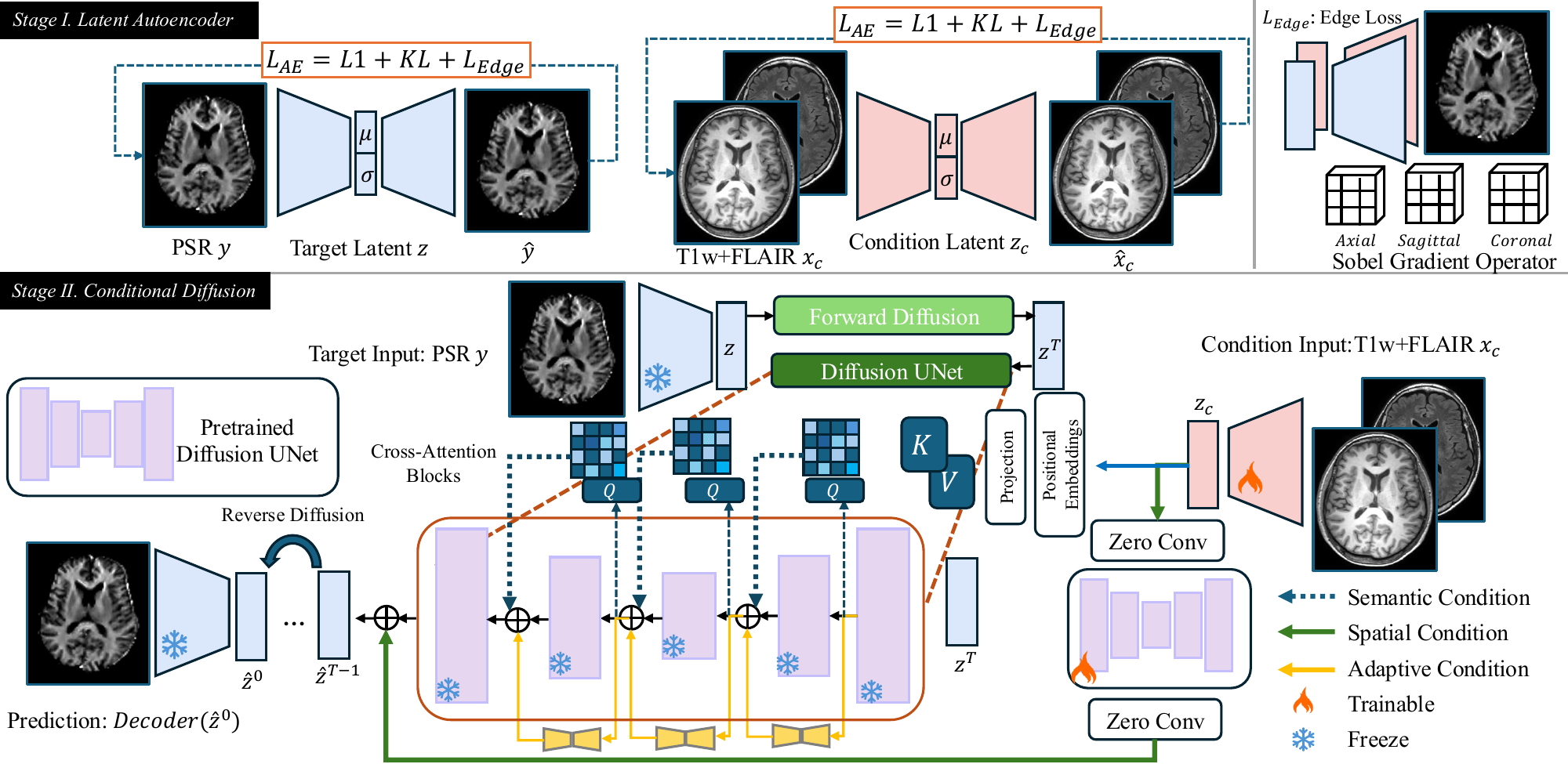}
    \caption{Overview of our framework. \textbf{Stage 1:} A latent diffusion model learns in PSR latent space. \textbf{Stage 2:} A pretrained diffusion backbone receives three complementary streams derived from \textcolor{BlueViolet}{\textbf{semantic}} tokens via cross-attention, \textcolor{ForestGreen}{\textbf{spatial}} residual hints from a 3D ControlNet, and \textcolor{Dandelion}{\textbf{adaptive}} LoRA on attention projections while keeping backbone weights fixed. \textbf{Edge-aware objective:} gradient-magnitude maps and from a 3D Sobel operator, sharpening generated brain tissue structures.}
    \label{fig:architecture}
\end{figure*}


\begin{itemize}
    \item \textbf{Novel framework for simultaneous anatomy and intensity conditioning:} Our goal is to preserve both anatomical structure and  PSR intensities. We achieve this with separate conditioning pathways in our diffusion model.
    
    \item \textbf{Data-efficient multi-scale conditioning:} We use separate autoencoders for each image type, apply ControlNet conditioning at multiple scales, and use 
    LoRA to adapt a frozen diffusion foundation model, avoiding degradation of pretrained priors. This design works well even with limited training data.

    \item \textbf{Clinically driven evaluation:} We include edge-preserving losses to maintain tissue boundaries, which are clinically important in MS. We validate our approach with standard image quality metrics as well as MS-specific metrics.
\end{itemize}


\section{Methods}
\label{sec:meth}

\noindent\underline{\textbf{Notation.}}
Let $x_{\mathrm{c}}\!\in\!\mathbb{R}^{2\times H\times W\times D}$ denote the  input images (T1w{+}FLAIR) and $y\!\in\!\mathbb{R}^{1\times H\times W\times D}$ the target PSR map. Two 3D AutoencoderKLs (same design, independent weights) define
$z_{\mathrm{c}}=E_{\mathrm{cond}}(x_{\mathrm{c}})$, $z_{\mathrm{y}}=E_{\mathrm{PSR}}(y)$, and $\widehat{y}=D_{\mathrm{PSR}}(z_{\mathrm{y}})$,
with spatial downsampling by $4\times$ so that any modality latent $z_{\mathrm{mod}}\!\in\!\mathbb{R}^{L\times \frac{H}{4}\times \frac{W}{4}\times \frac{D}{4}}$ (we use $L{=}8$).
We learn a conditional diffusion model entirely in the \emph{target} latent space to approximate $p(z_{\mathrm{y}}\mid z_{\mathrm{c}})$.
Diffusion follows the standard noise-prediction parameterization: at time $t$ the noisy latent $x_t$ is obtained by adding Gaussian noise $\varepsilon\!\sim\!\mathcal{N}(0,I)$ to $z_{\mathrm{y}}$, and the denoiser predicts $\varepsilon$ from $(x_t,t)$.

\noindent\underline{\textbf{Overview.}}
Our proposed DEMIST framework has two stages. Stage~1 learns separate autoencoders for PSR and T1w+FLAIR to obtain modality-specific latents that preserve salient structure. Stage~2 initializes a diffusion UNet from an unconditional BraTS-pretrained LDM \cite{antonelli2022medical,pinaya2023monai_generative,li2024brain} and keeps it frozen to retain the score prior. It then fuses anatomical conditioning into the denoiser through multi-level interfaces. This design separates anatomical alignment from quantitative mapping and enables data-efficient PSR synthesis.

We note that the unconditional 3D LDM pretrained on BraTS FLAIR volumes~\cite{pinaya2023monai_generative} in Stage 2 models $p(z)$ without conditioning, whereas our task is conditional image-translation in the PSR latent, learning $p(z_{\mathrm{y}}|z_{\mathrm{c}})$. Accordingly, we freeze the backbone to preserve the score prior and inject conditioning via  three decoupled streams: a \textcolor{BlueViolet}{ \textbf{semantic}} stream providing cross-attention tokens (what to synthesize), a \textcolor{ForestGreen}{\textbf{spatial}} stream using a zero-initialized 3D ControlNet to inject multi-scale residual hints at voxel level (where to place it), and an \textcolor{Dandelion}{\textbf{adaptive}} stream applying low-rank LoRA updates to selected attention projections (how to specialize the frozen prior to PSR with limited data). 

\subsection{Two-Stage Latent Diffusion}

\noindent\underline{\textbf{Stage~1: Latent Autoencoder Alignment.}}
We train two 3D KL autoencoders \cite{pinaya2023monai_generative} with the same architecture but independent (non-shared) weights: one for the target PSR domain $(E_{\text{PSR}},D_{\text{PSR}})$ and one for the conditional T1w{+}FLAIR domain $(E_{\text{cond}},D_{\text{cond}})$. Each encoder outputs mean
$\mu$
and standard deviation $\sigma$, defining a Gaussian posterior $q(z| \boldsymbol{\cdot})=\mathcal{N}(\mu,\mathrm{diag}\,\sigma^2)$. We reconstruct $\widehat{u}=D_{\boldsymbol{\cdot}}(E_{\boldsymbol{\cdot}}(u))$ for $u\!\in\!\{y,x_{\text{c}}\}$ and optimize:
\begin{equation}
\begin{aligned}
    \mathcal{L}_{\text{AE}}
&= \|u-\widehat{u}\|_1
+ \lambda_{\text{KL}}\,\text{KL}\!\big(q(z|u)\,\|\,\mathcal{N}(0,I)\big)\\
&+ \lambda_{\text{edge}}\,\|\nabla \widehat{u}-\nabla u\|_1
+ \lambda_{\text{adv}}\,\mathcal{L}_{\text{GAN}}(\widehat{u},u)
\end{aligned}
\end{equation}
Beyond the standard KL autoencoder objective, we add an edge-aware loss
$\mathcal{L}_{\text{edge}}=\|\nabla \widehat{u}-\nabla u\|_1$ based on first-order spatial gradients to encourage sharp tissue boundaries in reconstructions. 
After training both autoencoders, we freeze $E_{\text{PSR}}$ and fine-tune $E_{\text{cond}}$ to align the conditioning latent space with the target PSR latent space for Stage~2 diffusion process.

\noindent\underline{\textbf{Stage~2: Conditional Diffusion in the PSR Latent.}}
We adopt the standard LDM framework, defining a forward noising process on the clean PSR latent $z_{\text{y}}\!\sim\!q(\boldsymbol{\cdot}\mid y)$ with variance schedule $\{\alpha_t\}_{t=1}^{T}$ and cumulative product $\bar\alpha_t=\prod_{s\le t}\alpha_s$,
so that $z_t=\sqrt{\bar\alpha_t}\,z_{\text{y}}+\sqrt{1-\bar\alpha_t}\,\varepsilon$, where $\varepsilon\sim\mathcal{N}(0,I)$ and $z_t\in\mathbb{R}^{L\times h\times w\times d}$.
The denoiser $\epsilon_\theta(z_t,t;\mathcal{C}(z_{\text{c}}))$ predicts the added noise given the noisy latent $z_t$, timestep $t$, and  multi-level conditioning $\mathcal{C}(z_{\text{c}})$ derived from the conditional latent (detailed in Sec.~\ref{sec:condition}).
To leverage prior knowledge of brain MRI structure, we initialize the denoiser from a BraTS-pretrained unconditional 3D LDM \cite{pinaya2023monai_generative}. 
We freeze the pretrained backbone to preserve its score prior and only train $\mathcal{C}(z_{\text{c}})$ injected into the network. The training objective combines standard noise prediction with auxiliary terms that enforce edge preservation and latent alignment:
\begin{equation}
\begin{aligned}
\mathcal{L}_{\text{diff}}
&=\mathbb{E}_{t,\varepsilon}\!\left[\big\|\varepsilon-\epsilon_\theta\!\big(z_t, t;\,\mathcal{C}(z_{\mathrm{c}})\big)\big\|_2^2\right] \\
&+ \lambda_{\text{edge}}\;\big\|\nabla D_{\text{PSR}}(\widehat{z}_{\mathrm{y}})-\nabla y\big\|_1 
+ \lambda_{\text{align}}\;\big\|z_{\mathrm{c}}-z_{\mathrm{y}}\big\|_2^2,
\end{aligned}
\end{equation}
where \(\widehat{z}_{\mathrm{y}}=\big(z_t-\sqrt{1-\bar{\alpha}_t}\,\epsilon_\theta(z_t,t;\mathcal{C}(z_{\mathrm{c}}))\big)/\sqrt{\bar{\alpha}_t}\) denotes the predicted clean latent. The edge term encourages sharp tissue boundaries in the decoded output, while the alignment term regularizes the latent spaces to maintain consistent spatial correspondence between modalities.

\subsection{Decoupled Multi-Level Conditioning}
\label{sec:condition}
We construct $\mathcal{C}(z_{\text{c}})$ from the conditional latent through three complementary streams that address different aspects of the synthesis task: \textcolor{BlueViolet}{semantic} tokens (cross-attention), \textcolor{ForestGreen}{spatial} residuals (3D ControlNet), and parameter-efficient \textcolor{Dandelion}{adaptation} (LoRA) (Fig.~\ref{fig:architecture}). 

\noindent\underline{\textbf{\textcolor{BlueViolet}{Semantic stream.}}}
We project conditional latent $z_{\mathrm{c}}$ through a $1{\times}1{\times}1$ convolution, apply spatial pooling and add 3D sinusoidal positional encodings to obtain token sequence $C\!\in\!\mathbb{R}^{N\times d}$. Given the denoiser's spatially-flattened hidden feature $H\!\in\!\mathbb{R}^{M\times d}$ at each scale,
we apply standard dot-product cross-attention from $H$ (queries) to $C$ (keys $K$/values $V$) and add the residuals back to $H$.
This mechanism provides content-aware modulation that specifies what PSR characteristics to synthesize given the anatomical condition.

\noindent\underline{\textbf{\textcolor{ForestGreen}{Spatial stream.}}}
We incorporate ControlNet by creating a trainable copy of each frozen UNet block, initialized from the same BraTS-pretrained weights. The copy is connected to the frozen block through two zero-initialized $1{\times}1{\times}1$ “gate” convolutions: the first processes the conditional inputs $z_{\mathrm{c}}$ to match with frozen UNet input space and the second maps the output back to the frozen block's feature space.
The zero-initialization ensures the network initially behaves identically to the backbone.
During training, the trainable copy learns to output residual corrections that are added to the frozen block.
This provides stable, voxel-level spatial guidance (“where”) while keeping the main score network stable.

\noindent\underline{\textbf{\textcolor{Dandelion}{Adaptive stream.}}}
To enable data-efficient adaptation to PSR synthesis, we apply LoRA to all attention projection matrices ($W_Q,W_K,W_V,W_O$) at every resolution in the UNet. For each frozen weight matrix $W\!\in\!\mathbb{R}^{o\times i}$, we learn a low-rank update
$W \ \leftarrow\ W + \tfrac{\alpha}{r}\,BA$, where$
A\in\mathbb{R}^{r\times i},\; B\in\mathbb{R}^{o\times r},\; r\ll\min\{i,o\}$.
Only the low-rank factors $(A,B)$ are trainable while $W$ remains fixed. This design prevents overfitting on our limited training cohort while allowing the model to specialize from BraTS FLAIR domain to PSR.

\begin{figure*}[ht]
\centering
\begin{minipage}[t]{0.65\textwidth}
    \vspace{0pt} 
    \centering
    \resizebox{\linewidth}{!}{
        \begin{tabular}{lccc}
        \toprule
        Method & PSNR$\uparrow$ & SSIM$\uparrow$ & MSE ($\times 10^{-3}$) $\downarrow$ \\
        \midrule
        KL-GAN\cite{pinaya2023monai_generative} & 20.57 $\pm$ 3.52 &  0.8640 $\pm$ 0.0479 & 4.782 $\pm$ 5.680  \\
        VQ-VAE\cite{van2017neural} & 21.96 $\pm$ 3.79 &  0.8740 $\pm$ 0.0439 &  5.339 $\pm$ 8.252 \\
        FT-LDM\cite{rombach2022high} & 23.71 $\pm$ 3.34  &  0.8681 $\pm$ 0.0439 &  2.741 $\pm$ 8.624  \\
        cWDM\cite{friedrich2024cwdm} & 24.49 $\pm$ 2.87 & 0.8746 $\pm$ 0.0379 & 2.474 $\pm$ 8.077 \\
        \midrule
        FT-LDM + ControlNet & 24.52 $\pm$ 3.13 &  0.8741 $\pm$ 0.0419  & 2.317 $\pm$ 5.512  \\
        FT-LDM + LoRA & 25.16 $\pm$ 3.47 & 0.8811 $\pm$ 0.0442 &  2.012 $\pm$ 4.943\\
        \midrule
        \textbf{DEMIST (+ LoRA + ControlNet)} & \textbf{25.73 $\pm$ 3.56$^{*}$} &
        \textbf{0.8867 $\pm$ 0.0626$^{*}$} &
        \textbf{1.752 $\pm$ 3.878$^{*}$} \\
        \bottomrule
        \end{tabular}
    }
\end{minipage}%
\hfill
\begin{minipage}[t]{0.35\textwidth}
    \vspace{0pt} 
    \centering
    \includegraphics[width=0.95\linewidth,height=4.4cm]{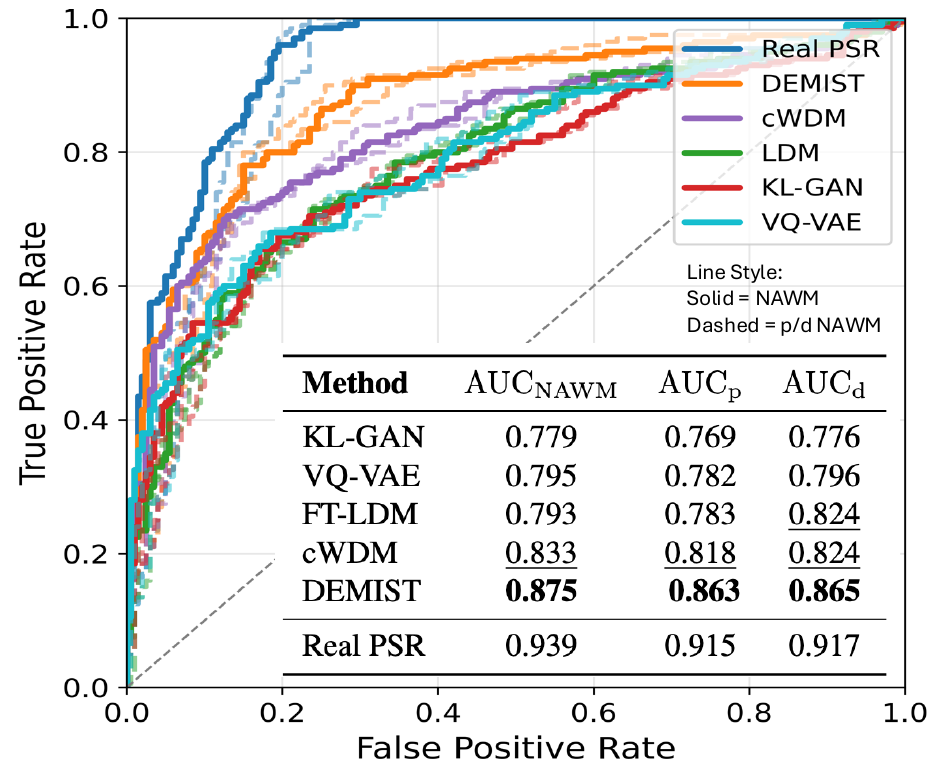}
\end{minipage}



{\captionsetup{skip=2pt}
\begin{minipage}[t]{0.38\textwidth}
    \captionsetup{type=table}%
    \caption{Cross-validation fold averages (5-fold CV over 163 scans). Best in \textbf{bold}. $^{*}$Statistically significant over the best baseline (cWDM).}
    \label{tab:main_results}
\end{minipage}%
\hfill
\begin{minipage}[t]{0.6\textwidth}
    \captionsetup{type=figure}%
    \caption{ROC curves discriminating T2 lesions from NAWM using PSR values. Colors indicate different methods (\textcolor{blue}{blue}: ground truth, \textcolor{orange}{orange}: ours). Solid lines: combined NAWM; dashed lines: proximal and distal NAWM.}
    \label{fig:roc}
\end{minipage}
}
\end{figure*}


\begin{figure*}[t]
    \centering
    \includegraphics[width=\linewidth]{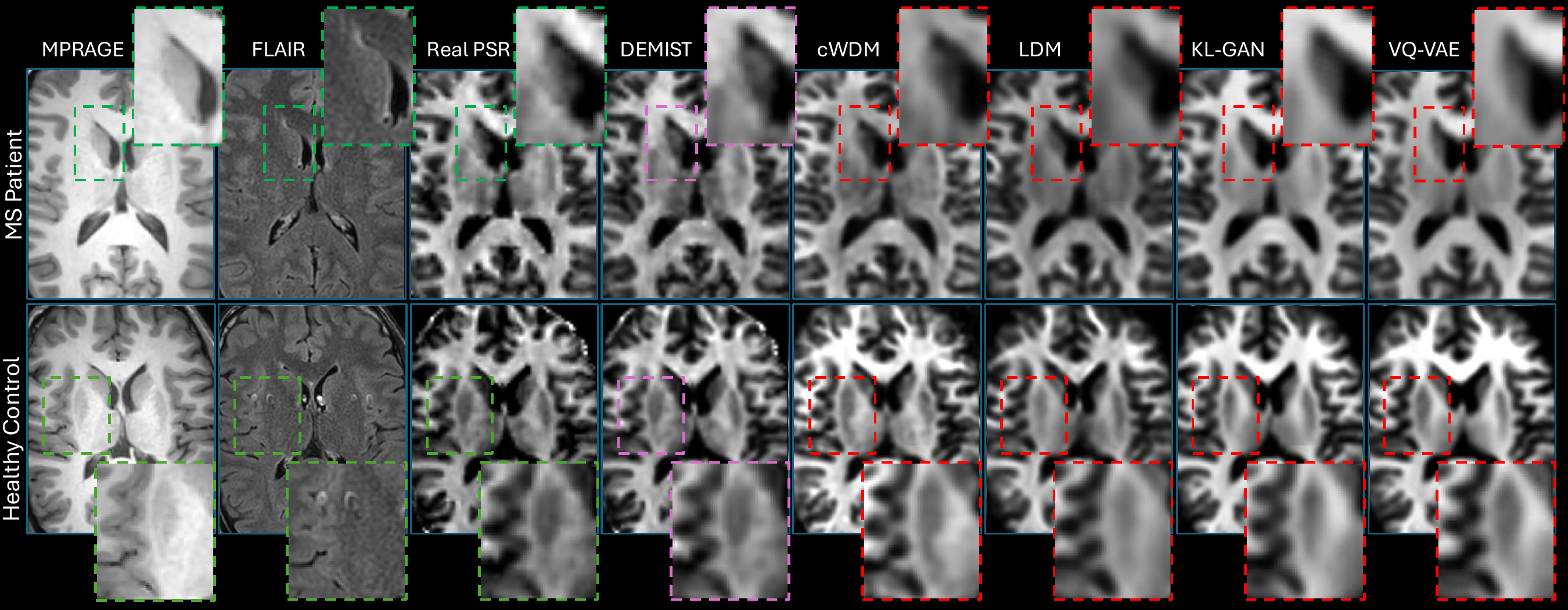}
    \caption{Qualitative results. Each row shows conditions (\textcolor{ForestGreen}{MPRAGE}, \textcolor{ForestGreen}{FLAIR}), ground truth \textcolor{ForestGreen}{PSR}, and synthesized outputs from \textcolor{pink}{our method} and the \textcolor{red}{baseline} methods. 
    Zoom panels highlight our method produces sharper boundaries for the lesion (in top row, an MS patient), and provides more accurate anatomy (highlighted in the putamen shape in the bottom row, a healthy control).}
    \label{fig:qualitative}
\end{figure*}

\section{Results and Discussion}
\label{sec:experiments}

\noindent\textbf{\underline{Dataset and Preprocessing:}}
We evaluate our method on 163 3T MRI scans from 99 subjects (68 MS patients and 31 healthy controls) acquired at a VA-affiliated site and Vanderbilt University Medical Center. Subjects were scanned longitudinally with 1-3 timepoints per patient (mean 1.67 $\pm$ 0.78). Each scan includes T1-weighted MPRAGE, T2-weighted FLAIR, and qMT-derived PSR maps.

To prevent data leakage between timepoints, we use 
5-fold cross-validation where all scans from the same patient are assigned 
to the same fold. The train/validation split for each fold is approximately 130/33 scans.
All images are co-registered, reoriented to RAS, and resampled to 1mm$^3$ isotropic resolution. We normalize intensity values to $[0,1]$ using 0-99.5th percentile scaling. For training, volumes are 
divided into overlapping patches $96\times96\times96$. Training details can be found at \url{MedICL-VU/MS-Synthesis-3DcLDM}.

\noindent\textbf{\underline{Evaluation Protocol:}}
We assess image reconstruction fidelity using peak signal-to-noise ratio (PSNR), structural similarity index (SSIM), and mean squared error (MSE). All metrics are computed within the brain.

To evaluate whether synthesized PSR preserves differences between pathological and healthy tissue, we also report the area under the receiver operating characteristic curve (AUC) for discriminating T2 lesions from normal-appearing white matter (NAWM). Following the validation approach of Toubasi et al.\ \cite{toubasi2025improving}, we analyze 231 lesion-NAWM pairs from the first scan of 40 MS patients, comparing both proximal NAWM (pNAWM, adjacent to lesions) and distal NAWM (dNAWM, contralateral regions) against lesion voxels.

\noindent\textbf{\underline{Baselines:}}
We compare our model against (i) \textbf{KLGAN} \cite{song2020bridging} , (ii) \textbf{VQ-VAE} \cite{van2017neural}, (iii) \textbf{FT-LDM} \cite{rombach2022high} (fine-tuning pretrained LDM \cite{pinaya2023monai_generative} without ControlNet residuals or LoRA), and (iv) \textbf{cWDM} \cite{friedrich2024cwdm} (best performance in BraSYN challenge \cite{li2024brain} using wavelet transform). For fairness, all methods are trained on the same 5-fold splits with the same preprocessing.


\noindent\textbf{\underline{Image-level metrics.}}
Table~\ref{tab:main_results} reports test-fold results (mean$\pm$SD across folds). Our model achieves best performance across all metrics, including against the state-of-the-art cWDM as well as the fine-tuned pretrained LDM baseline (FT-LDM).  Our model also shows better structural similarity (SSIM) and voxel-wise error (MSE), as well as L1 and NCC metrics (not shown for brevity). The improvement is especially pronounced in the error metrics (MSE, L1). All improvements were statistically significant ($p<0.05$ in paired t-tests).

\noindent\textbf{\underline{Ablation Studies.}}
We also report an ablation study to isolate contributions of ControlNet and LoRA. Table~\ref{tab:main_results} shows LoRA provides the largest improvement (PSNR, SSIM, MSE), demonstrating that low-rank adaptation is critical for data-efficient PSR learning. ControlNet yields smaller but consistent gains in structural alignment and spatial consistency. Its multi-scale residual stabilizes voxel-level consistency. 

\noindent \textbf{\underline{Tissue contrast preservation.}}
Figure~\ref{fig:roc} shows our method achieves the best AUC values for combined NAWM, for pNAWM, and for dNAWM. This performance gets substantially closer than any compared methods to the measured PSR performance (oracle). The consistent performance across proximal and distal regions indicates that our model captures spatially varying tissue characteristics rather than learning average intensity differences alone.

\noindent\textbf{\underline{Qualitative Analysis.}}
Figure \ref{fig:qualitative} shows representative results for both MS patients and healthy controls. Our method produces sharper lesion boundaries and better preserves cortical gray-white matter transitions. The GAN-based and VQ-VAE approaches tend to blur fine anatomical details, while the standard LDM and cWDM methods provide fine details but underestimate PSR contrast.

\noindent\textbf{\underline{Discussion.}}
We proposed DEMIST: a 3D latent diffusion mode to synthesize qMT PSR maps from standard T1w and FLAIR images using frozen pretrained weights with ControlNet and LoRA adaptation. Our method outperforms GAN and diffusion baselines, enabling PSR synthesis without acquiring qMT.
Future work will validate on data from multiple sites and scanners. 
Inference is currently relatively slow due to iterative sampling.
In future work, we plan to explore distillation methods for faster sampling, and investigate using user-provided hints (like lesion masks) for better control.



\clearpage

\noindent\textbf{Acknowledgments.}
This work was supported, in part, by the U.S. Department of Veterans Affairs (I01CX002160-01A1), the National Multiple Sclerosis Society (RG-1901-33190) and Voros Innovation Impact Funds. The authors declare no conflicts of interest.

\noindent\textbf{Compliance with Ethical Standards.}
This retrospective study was approved by the Vanderbilt University Medical Center Institutional Review Board. All patient data were de-identified prior to analysis in accordance with HIPAA regulations.

\bibliographystyle{IEEEbib}
\bibliography{refs.bib}

\end{document}